\documentstyle[12pt,preprint,revtex]{aps}
\begin{document}


\begin{title}
Polarons in the Three-Band  Peierls-Hubbard Model: An Exact
Diagonalization Study
\end{title}

\author{A.Dobry$^*$}
\begin{instit}
Groupe de Physique Statistique, Universit\'{e} de Cergy Pontoise\\
47 rue des Genottes,95806 Cergy Pontoise Cedex\\
 France
\end{instit}
\author{A.Greco$^*$}
\begin{instit}
Max-Planck Institut f\"ur Festk\"orpersforschung\\
Heisenbergstr.~1, D-7000 Stuttgart 80\\
Germany
\end{instit}
\author{J.Lorenzana}
\begin{instit}
Laboratory of Applied and Solid State Physics. University of Groningen\\
Nijenborgh 4, 9747 AG Groningen\\
The Netherlands
\end{instit}
\author{J.Riera}
\begin{instit}
Physics Division, Bldg.6003, Oak Ridge National Laboratory\\
Oak Ridge, TN 37831
\end{instit}

\begin{abstract}

We have studied the three-band Peierls-Hubbard model describing the
Cu-O layers
in high-T$_c$ superconductors by
using Lanczos diagonalization and assuming infinite mass for the ions.
When the system is doped with one hole, and for
$\lambda$ (the electron-lattice coupling strength ) greater
than a critical value, we found that
the oxygens around one Cu contract and the hole self-traps
forming a lattice and electronic small polaron.
The self-trapped hole forms
a local singlet analogous to the Zhang-Rice singlet in the undeformed
lattice.
We also studied the single-particle spectral
function and the optical conductivity.
We have found that the spectral weight, in general, is  similar
to that found in
previous studies where the coupling with the lattice was absent.
There is an anomalous transfer of spectral weight but, contrary to those
studies, the spectral weight goes to these localized polaronic states.
However this polaronic shift does not seem enough by itself
to  explain pinning of the chemical potential observed in real
materials.
The peaks in the optical
conductivity are also shifted, according to
the polaronic shift, in the single-particle spectral functions.
We compare our results to those obtained in inhomogeneous Hartree-Fock
calculations and we discuss their relation with experiments.

\end{abstract}


\newpage

\section{Introduction}

\indent

Since the discovery of high-T$_c$ superconductors\cite{1},
a considerable effort has  been made to determine the
nature of hole-doped states in the copper-oxide planes which are
present in these materials.
Many theoretical
discussions have emphasized the importance of strong electronic
correlations within these planes, leading to the study of purely electronic
two-dimensional models like the one-band Hubbard model \cite{2}, the
$t-J$ model \cite{3} and three-band Hubbard models \cite{4}.
In spite of the fact that the studies on these purely electronic
models are still inconclusive,
there are clear experimental indications that, at least in the dilute
regime, the coupling  to the lattice should be taken into account \cite{latt}.

The nature of doping states with and without coupling to the lattice
has been studied in inhomogeneous
Hartree-Fock (HF) calculations for one-band and multi-band Hubbard
models\cite{lor2,lor3,inu,8,7,14,13,17}.
In the three-band Hubbard model, with no coupling to the lattice,
a doped hole
localizes mainly in the four surrounding O sites of
a Cu,
nearly canceling the spin density at the central Cu\cite{8,7,14,13,17}.
This charge and magnetic polaron is a manifestation, at the HF level, of
the singlet bound state that
Zhang and Rice\cite{3} found in  their
derivation of the $t-J$ model (as an effective Hamiltonian of the
strongly interacting three-band Hubbard model).
By including the coupling with the lattice \cite{8,17}
it was found that
the O's around the central Cu forming the electronic polaron relax
with a  symmetric
pattern [Fig. 1(b)]. Similar ideas were proposed earlier in
Ref.~\cite{ric}. The infrared (IR) spectra of this state\cite{8,17},
for moderate electron-lattice coupling, calculated in the
random phase approximation (RPA), shows a doped-induced
bleaching of the in-plane Cu-O stretching  mode and a side phonon
band at lower frequencies.
The nature of the charge carriers and their interactions
with the lattice has been probed by infrared optical
absorption in photodoped\cite{9,10,gyu} and chemically doped\cite{11}
samples.
Both kind of experiments show in La$_2$CuO$_4$, among other materials,
a doping-induced bleaching of phonon modes and, in general,
an intensity shift to
lower frequencies. Of particular interest in La$_2$CuO$_4$  is the
bleaching of the 708 cm$^{-1}$ Cu-O
stretching mode and the appearance  of a side band at
640 cm$^{-1}$\cite{9,gyu}. Both are
clearly resolved in the photodoped spectra and consistent with the
data of chemically doped samples
\cite{11}. This behavior agrees wells with the HF plus RPA
calculations and provides a strong indication of the existence of
the electronic and lattice polarons.

In spite of this success of HF plus RPA calculations, the above studies
present various shortcomings.
Since
the  electronic polaron is already self-trapped {\em without}
coupling to the lattice, a negligible small electron lattice coupling
strength  $\lambda$
(defined below)  suffices to produce
lattice relaxation. In other words, there is no  competition between
the delocalization energy and the self-trapping effects  in
this type of calculations.
Another serious shortcoming comes from the fact that the Zhang-Rice
wave function
can not be represented by a single Slater determinant, and ( at least at
the HF level) a relevant part of the physics is missing.
In this sense, it is important to
study the interplay between electronic correlations and lattice defects
or deformations
without the above limitations.
The electronic correlations can stabilize
the lattice deformations as a consequence of the pre-existing
enhancement of the effective mass.
This effect has been observed in an
exact diagonalization study of the Holstein-Hubbard model\cite{15}.
On the other hand, the lattice deformations can
introduce anharmonic effects, further enhancing the effective mass
and  changing the properties of carriers
with respect to those  found in the
framework of purely electronic models.

In this paper we study, by means of Lanczos diagonalization, the formation
and properties of polarons in the two dimensional(2D) three-band
Peierls-Hubbard model, which we believe
is a realistic microscopic model of the Cu-O planes.
The model is presented in Section II.
We take the mass of the O ions equal to infinity (adiabatic
approximation) for simplicity.
This method is free from the above mentioned limitations of HF
approximations and takes into account all the advantages of
exact diagonalization studies of the
three-band Hubbard models\cite{16,P.Hors} which
have reproduced fairly well the main expected  spectral signatures.
We discuss finite size effects in the conclusions (Section VII).
By minimizing the total energy, we study the range of electron-lattice
coupling at which the  various lattice deformations become stable.

For any finite electron lattice coupling strength $\lambda$ below
a critical value ($\lambda_c$), we find that the system forms a
charge density wave (CDW) with a
breathing extended lattice distortion (BED) [Fig. 1(a)]. A similar distortion
was found {\em above} a critical value in Monte Carlo simulations \cite{mur}
of the Hubbard model. In Section III we discuss the phase diagram in the
parameters space
and give general arguments of why a BED should be found for any finite
$\lambda$.

Above $\lambda_{c}$ a polaron is form.
In what follows, we use the term polaron to designate the composite
object made by the lattice deformation and the self-trapped carrier.
The polaron results from static oxygen displacements as shown in
Fig. 1 (b).
This is in accordance with the HF picture, but
in exact diagonalization calculations, the polaron appears only
above $\lambda_c$ whereas, as explained before, in HF it appears for any
finite $\lambda$.

The polaron and the CDW are  characterized by
studying static magnetic and density-density correlation functions
(Section IV). The presence of the self-trapped carrier manifests also in
the single-particle spectral function. The
evolution of the states formed by these self-trapped carriers as a function
of $\lambda$ is study in Section V. Special
attention is given
to the issue of anomalous transfer of spectral weight
and evolution of the chemical potential with doping\cite{22,23}, in
view of the controversial experimental situation\cite{vee}.
Finally, the effect of the self-trapped states on the optical conductivity
is analyzed in Section VI. Consistently with the features of the
single-particle spectra discussed above, we find a shift of the peaks
to higher energies.

\section{The Model, Parameters and Cluster Studied}

\indent

We consider the following Hamiltonian for the Cu-O planes:
\begin{eqnarray}
H = \sum_{\scriptstyle i \neq j , \sigma} t_{ij}(\{u_{k}\})
c^{\dagger}_{i,\sigma}
c_{j,\sigma} + \sum_{\scriptstyle i,\sigma} e_{i} c^{\dagger}_{i,\sigma}
c_{i,\sigma}\nonumber\\
+ \sum_{i} U_{i} n_{i,\uparrow}n_{i,\downarrow}
+ \sum_{i \neq j} U_{ij} n_{i} n_{j}
\nonumber\\
+\sum_{j}\frac{1}{2}K u_j^{2}
\label{uno}
\end{eqnarray}

Here, $c_{i,\sigma}^{\dagger}$ creates a hole with spin $\sigma$ at the
site
$i$ in the Cu $d_{x^2-y^2}$ or the O $p_{x,y}$ orbitals,
$n_{i,\sigma} = c^{\dagger}_{i,\sigma} c_{i,\sigma}$ and
$n_{i} = (n_{i,\uparrow} + n_{i,\downarrow})$.
We have included
on-site repulsions $U_{i}=U_d,U_p$ for holes on Cu and O atoms and a
 nearest-neighbor
Cu-O repulsion $U_{ij}=U_{pd}$. $e_{i}=e_d,e_p$ are the site energies of
the Cu and O atoms.

The static lattice deformation and the electron-lattice coupling are
considered in the following way.
Each of the patterns of deformation considered here (see next section) are
defined by a set of displacements $u_{j}$ of the oxygen ions along the
Cu-O bonds.
We assume that the nearest-neighbor Cu-O hoppings are modified by the O-
ion displacements as $t_{ij} = t_{pd}\pm \alpha u_{j}$, where the + (-)
applies if the bond shrinks (stretches) with positive $u_j$.
In order to study the stability of the system with respect to a given
deformation of the lattice, we included the elastic energy in the
Hamiltonian (last term of Eq. \ref{uno}).
For a given value of $\lambda$, and for a set of displacements $u_j$,
we diagonalize exactly the electronic part of the Hamiltonian using
the Lanczos algorithm. Then the total energy is minimized by
varying $ \mid u_j \mid$ constrained to the specific pattern.

As a reference parameter set we take
$\Delta=e_p-e_d=3,U_d=8,U_p=3,U_{pd}=1,t_{pd}=1$.
For simplicity the O-O hoping $t_{pp}$ is taken equal to zero.
We take infinite masses for the O ions,
 so that their kinetic energy drops out.
Following the estimates made in Ref.~\cite{8,17} we take
$K=32 t_{pd} \AA^{-2}$.
In terms of the above parameters, we have defined the
electron-lattice coupling
as $\lambda=\frac{\alpha^2}{Kt_{pd}}$.

The exact diagonalization study was made on the Cu$_{4}$O$_{8}$ cluster
(12-sites) with periodic boundary conditions. We consider two cases,
the half-
filled system (one hole per Cu-O$_2$ cell), and the system doped with
one extra hole.

\section{Study of the Stability of Various Lattice Deformations}

\indent

As a first step in the analysis of the electron-lattice coupling, in
this section we study
the stability of an homogeneous
distortion of the lattice (BED)
[Fig.~\ref{fig1}(a)] and the stability of
a local contraction of four oxygens toward
one Cu (breathing ``localized" deformation or BLD)
[Fig.~\ref{fig1}(b)]. We shall also refer to this localized deformation
as lattice polaron.

In the half-filling condition the lattice remains stable upon  these
breathing deformations, for all physically reasonable
values of $\lambda$ ($\lambda \tilde{<}  7$). A similar result was found
in a Monte-Carlo study of the analogous one-band model\cite{mur} and in
the HF calculations \cite{8,17}.

In general, the system can accept a lattice deformation if it gains enough
local covalency to compensate the cost of elastic energy,
delocalization energy,
magnetic energy or Coulomb repulsion.
At half-filling, the delocalization energy is obviously not relevant.
We should analyze only the magnetic energy, Coulomb repulsion and covalency.
Without deformation, the spin density at each Cu site is uniform
and equal to a fixed value, say $m$. When  the oxygens are displaced, the
hopping integrals are modified and it is likely for  charges to
spend more time around the Cu sites of one sublattice and less time in the
Cu sites of the other sublattice.
Thus, for both Cu sites, the spin density $m$ is
reduced because, in one case, there is an increased tendency to double
occupation, and in the other, the sites are becoming increasingly empty.
The main increase in the energy (of order $U_d$) comes from the double
occupancy and a large value of $\lambda$ is necessary to compensate it.
More complicated patterns of deformation, not studied here,
can gain some
covalency without increasing the double occupancy too much\cite{13,8}.
and, presumably,  can destabilize  the lattice for smaller $\lambda$.

The situation changes upon doping.
With one doped hole, the BED is stable for every nonzero
$\lambda$ smaller than a certain
critical value $\lambda_{c}$ ( $\lambda_{c} \approx 1$ for our set of
parameters). The fact that the BED is stable for  arbitrary small
$\lambda$  can be understood as follows :
the ground state of the undeformed system is
four-fold degenerated with momentum
$k=(\pi,0)$ or  $k=(0,\pi)$ and $S_z=\pm 1/2$. Consider now the
non-interacting case ($U_p=U_d=U_{pd}=0$) with the same number
of particles.
The ground state is formed by
five-particle Slater determinants which, for fixed  $S_z$, can be
labeled by the total momentum as,
$|(\pi,0)>$ or $|(0,\pi)>$. We can construct two states
$|+>=\frac{1}{\sqrt{2}}(|(\pi,0)>+|(0,\pi)>)$ and
$|->=\frac{1}{\sqrt{2}}(|(\pi,0)>-|(0,\pi)>)$. Taking the $x$ coordinate
in the direction of the Cu$_1$-Cu$_2$ bond in Fig.~\ref{fig1}
it is easy to see
that $|+>$ will have more charge in sites 1 and 4 and $|->$ in sites 2
and 3. Clearly the BED distortion favors one of these configurations at
the expense of the other. In other words, the BED lowers the symmetry
of the lattice lifting the degeneracy.
This conclusion is valid for the interacting case
because the ground state in this system
has the same symmetry and quantum numbers as in the noninteracting one.

For one doped hole, we show in
Fig.~\ref{fig2} the ground state energy, measured with respect to the
energy of the undistorted lattice, as a function of $\lambda$.
Both localized and extended deformations are shown.

For $\lambda > \lambda_{c}$, the BLD becomes stable.
The loss in elastic, Coulomb
and delocalization energies is compensated by the gain in local covalency.
In order to understand this crossover, we show in Fig.~\ref{fig3},
the equilibrium lattice displacements of the oxygens along the
Cu-Cu axis as a function of $\lambda$ for both types of deformations.
In the region $\lambda < \lambda_{c}$ the displacements of the ions
are very small and, consequently, there are little
effects of the deformation on the ground state properties.
When $\lambda$ becomes larger than $\lambda_{c}$ the four displaced
oxygens of the BLD undergo a sudden contraction toward their
central copper. At this point this localized deformation becomes
stable and the extra hole is trapped in this Cu-O$_{4}$
cluster, forming a small polaron state.

We also studied the variation of $\lambda_{c}$ when $U_d$ is
moved away from the reference parameter set (see Fig.~\ref{fig4}).
It can be seen that $\lambda_{c}$ increases as a function of $U_{d}$
roughly linearly in the studied range ($ 6 \leq U_{d} \leq 20 $).
This  is because  BLD becomes stable by
localizing the added hole (see next section) which, in turn, leads to an
increasing double occupancy of the central Cu site costing an
energy of order $U_d$. Thus, it is necessary a larger electron-
lattice coupling to overcome this Coulomb repulsion.

It is worth noting that, in the strong coupling limit,
this modification of the hopping integrals due to deformation
leads also to a local
reduction in the effective antiferromagnetic coupling $J$.
In
fact, in a perturbative calculation, $J$ results proportional to the
fourth power of the hopping
integral $t$ and then, upon deformation
$J \approx (t^{2}-(\alpha u)^{2})^{2} < t^{4}$. The loss in magnetic
energy,
due to the lattice deformation
discussed above, could also be related to this reduction in $J$.

\section{Static correlation functions}

\indent

The formation of the polaron when one  hole is added to the system can be
determined by analyzing, the spin and charge densities.

In Fig.~\ref{fig5}  we show the sum of the occupancies $<n_i>$ in the
central Cu (Cu$_1$) and its four surrounding O involved
in the BLD, as a function of $\lambda$. The total occupancy of
this Cu-O$_4$ cluster is $\approx 2$ for $\lambda> 1$
indicating the trapping of the extra hole.
We also show the occupancy of the Cu$_4$ site, the most distant Cu from Cu$_1$,
and
its four surrounding O.

In Fig.~\ref{fig6} we display $<S_z^2>$ in Cu$_1$, Cu$_4$, and in the two
remaining Cu sites (Cu$_2$ and Cu$_3$),
as a function of $\lambda$.
The reduction of the magnetic moment at the central Cu results
from the fact that the two holes, with opposite spins, are
jumping back and forth between the central copper and its four
surrounding oxygens. As a consequence of this there is an increased
tendency to double occupancy in the central Cu.
The reduction of the
magnetic moment at the central Cu was also, found with qualitatively
similar behavior, in Ref. \cite{8,17}.

The  coupling with the lattice
takes advantage of a preexisting enhanced effective mass, and
produces the self-trapping of a Zhang-Rice
singlet. To follow the evolution of this singlet we define the
following operator:
$Z_{i}^{\dagger}= (P_{i,\uparrow}^{\dagger} d_{i,\downarrow}^{\dagger} -
P_{i,\downarrow}^{\dagger} d_{i,\uparrow}^{\dagger})$ where
$P_{i,\sigma}^{\dagger}$ creates a hole with spin $\sigma$ in a symmetric
combination of the four oxygen orbitals. In Fig.~\ref{fig7} we
show the expectation
value of $<Z^{\dagger}Z>$ in the ground state as a function of $\lambda$.
This quantity is related to the probability of finding a ZR state in the
cell. We can see a rapid increases of this value when the polaron
is formed.
However,  the double occupancy also increases  at the
central Cu. The amount of double occupancy could be measured from its
order parameter $<n_{i,\uparrow} n_{i,\downarrow}>$ computed at
Cu$_1$. We found $25\%$
of double occupancy when the polaron is formed, while in the undeformed
lattice this value is only $6\%$. From this, we conclude that the
polaron is  a ZR  singlet but of a more general character that
the one found
in the undeformed lattice.

\section{Single Particle Spectral Function}.

\indent

In this Section, we analyze the evolution of
the spectral weight for single-particle
excitations when the coupling to the lattice is switched on.
We first review the previous known results\cite{16,P.Hors,21,oht},
for $\lambda=0$.
There has been some theoretical and experimental controversy about
how the Fermi level and the spectral
weight evolve with doping in real materials and therefore
we pay special attention
to these issues, in the light of the present results. In particular, some
experiments\cite{24} suggest that the chemical potential remains pinned
in the gap of the insulator as the system is doped, whereas
others\cite{vee},
suggest that it shifts
with doping as in an ordinary semiconductor. In both cases the gap tends
to fill in with transferred spectral weight.

The single-particle spectral function is defined as,

\begin{eqnarray}
g_{i\sigma}\left(\omega\right) = g_{i\sigma}^> \left(\omega\right) +
g_{i\sigma}^<\left(\omega\right)
\label{dos}
\end{eqnarray}

where

\begin{eqnarray}
g_{i\sigma}^>\left(\omega\right) = \sum_n |<\psi_n\left(N+1\right)|
c_{i\sigma}^\dagger|\psi_0\left(N\right)>|^2 \times \delta\left(\omega-
E_0\left(N\right)+E_n\left(N+1\right)\right).
\label{tres}
\end{eqnarray}

To obtain $g_{i\sigma}^<\left(\omega\right)$ we have to change $c_{i\sigma}
^\dagger$ and $N+1$ by $c_{i\sigma}$ and $N-1$ respectively. $|\psi_n\left(N
\right)>$
is an eigenstate having eigenvalue $E_n\left(N\right)$ where $N$ is the
number of holes; $E_0(N)$ is the ground state energy. All the energies
and gaps given below are in units of $t$.
$g^>$ and $g^<$
describe the photoemission (PES) and inverse photoemission (IPES)
spectra respectively.

Following Refs. \cite{16} and \cite{P.Hors},
the expression for $g_{i\sigma}\left(\omega\right)$
can be evaluated numerically using the Lanczos algorithm and the continued
fraction expansion formalism\cite{19}.

In Fig.~\ref{gstoc}
we show the single-particle spectral function
($g_i=g_{i\uparrow}+g_{i\downarrow}$)
at the Cu and O sites in the undoped (half-filling) case
$\lambda = 0$, i.e. for the undeformed lattice.
The main features of this spectra have been identified in a previous
study\cite{16}.
First of all, we can distinguish
the lower (LHB) and upper (UHB) Hubbard bands, which are well separated
by $U_d$ plus a typical hybridization energy. Below the UHB we find the
nonbonding oxygen band (NOB) with a small weight on the Cu sites. The energy
difference between these two bands is slightly larger than
$\Delta + 2U_{pd}$. The quantity
$\Delta + 2U_{pd}$ is the (bare) charge-transfer gap expected from
a ionic limit.

Between the UHB and the NOB other structure emerges. We have indicated the
momentum of the states close to the  Fermi energy.
All of them have total spin 1/2. Due to the spatial symmetry of this lattice,
the peak
with momentum $k=(0,0)$ is absent\cite{20} but it reappears if
$t_{pp}$ is switched on\cite{21}. These highly
correlated states (CS)  correspond to the Zhang-Rice singlets band\cite{16}.
The gap is given by the
energy difference between UHB and CS. This energy difference is $\sim 3 $
for  our parameter set.

When the system is doped with one hole and $\lambda=0$ (Fig.~\ref{g0})
the Fermi level is shifted to $\mu^+=E_0(5)-E_0(4)$\cite{22,23}.  This is
exactly the position of the $k=(\pi,0)$ peak in Fig.~\ref{gstoc}.
This can be contrasted with the HF mean field calculations\cite{14} where
the stoichiometric system does not show
spectral weight in the  place where the chemical potential of the
one hole doped systems sits. This is worth  emphasizing  because
it shows that the HF results cannot be automatically invoked as an explanation
of the photoemission experiments showing pinning of the chemical
potential\cite{24}.
The proximity of the next available state in Fig.~\ref{g0} indicates
that the system is metallic.
The low energy spectral weight (LESW) between the Fermi
level and the gap in Fig.~\ref{g0},
as defined in Ref. \cite{22,23}, is equal to 1.85. This
quantity should be equal to 1 for an ordinary semiconductor,
and equal to
2 for the Hubbard model in the atomic limit. The value close to 2 is in
agreement with the results of Ref. \cite{22,23}  and it is due to
strong covalency.
It is worth noting that, at the HF level,
departures from an integer value of LESW (in this case 2) are not allowed.
Transitions from these states to
scattering   states above  the Fermi level were
identified\cite{14} as the origin of the mid infrared absorption
in the optical conductivity.

For $0<\lambda<\lambda_c$, as discussed
in the previous sections, the symmetry of the lattice is lowered and
hence
the charge tends to accumulate around the Cu$_1$ and  Cu$_4$ sites (Fig. 1).
This produces
changes in the local $g$'s, but the total spectral weight remains
similar to the undistorted one and  the system stays   metallic.

As we have shown in Section III,
the BLD becomes stable
in the region $\lambda > \lambda_c$.
In this
case, the translational invariance is broken further.
In Fig.~\ref{g1.2}, we show the spectral weight for $\lambda=1.2$.
The self-trapping of the carrier around Cu$_1$ makes the local
hole-occupied $g$ to increase. The same quantity is larger in
Cu$_4$ than in Cu$_2$ indicating that there is still some amplitude
for the CS state to hop out and in from the Cu$_1$,  mainly to the
next nearest neighbor's cell. The local $g$'s for the oxygens (not shown)
around Cu$_1$ (Cu$_4$) increase the hole-occupied (-unoccupied)
spectral weight close to the Fermi level. Apart for this feature, they
are similar to the total oxygen $g$ [Fig.~\ref{g1.2}(e)].

The levels located slightly above the Fermi level in Fig.~\ref{g1.2}
show a polaronic shift. This makes the $\mu^+$
to sit, not in the position of the first peak in Fig.~\ref{gstoc}, like
in the $\lambda=0 $ case, but somewhere in the gap. This is a direct
consequence of the fact that we took an infinite mass for the O
ions. If the mass were finite
we expect Fig.~\ref{gstoc} to be $\lambda$-{\em dependent}.
Accordingly, some intensity will show up  at the position of the new $\mu^+$
corresponding to the polaronic configurations. However, for an infinite
mass, such configurations
can not be reached quantum mechanically by adding one particle to the
half-filled case. Thus,  their spectral weight vanish.
It is tempting to associate this behavior with the experimental
observation of growing spectral intensity and pinning of the chemical
potential in the gap upon
doping\cite{24}. The  small spectral intensity in the undoped case
due to a finite O mass will be hardly measurable
and hence interpreted as the gap. Nevertheless  the polaronic shift close
to $\lambda_c$ is too small by itself to explain the effect.
It can help to this interpretation when combined with
other effects like impurities.
The low energy spectral weight in this case is 2.06. This increase
is expected from a higher  covalency but, in our case, contrary to
Ref. \cite{22,23}, this effect is a local one.  For even larger values
of $\lambda$ the polaronic shift increases further in qualitative
agreement with Ref. \cite{8,17}.

\section{Optical Absorption}

\indent

In this Section, we report our results for the optical conductivity.
The calculation of this quantity in the framework of exact
diagonalization follows the procedure explained in detail in
Refs. \cite{16,19}
The optical absorption $\sigma\prime
\left(\omega\right)$ or the real part of the complex conductivity
$\sigma\left(\omega\right)$ is defined as:

\begin{eqnarray}
\sigma\left(\omega\right)=\frac{1}{i\omega}[\frac{\tau^2}{N_{sites}}<
0|T_y|0>
-\chi\left(\omega\right)]
\label{cuatro}
\end{eqnarray}

where

\begin{eqnarray}
\chi\left(\omega\right)= \frac{1}{N_{site}}\sum_n|<0|j_p|n>|^2 [\frac{1}{
\omega-\left(E_n-E_0\right) +i\eta} - \frac{1}{\omega+\left(E_n-E_0\right)
+i\eta}].
\label{cinco}
\end{eqnarray}

The quantities:

\begin{eqnarray}
j_p=i\tau\sum_{i\sigma}t_{ii+y}(\{u_k\})\left(c_{i\sigma}^\dagger
c_{i+y\tau,\sigma}
- c_{i+y\sigma}^\dagger c_{i\tau,\sigma}\right)
\label{seis}
\end{eqnarray}

and

\begin{eqnarray}
T_y=\sum_{i\sigma}t_{ii+y}(\{u_k\})\left(c_{i\sigma}^\dagger c_{i+y\tau,\sigma}
+ c_{i+y\sigma}^\dagger c_{i\tau,\sigma}\right),
\label{siete}
\end{eqnarray}

\noindent
are the current and kinetic energy in the direction of the applied field,
respectively. $\tau$ is the distance between two nearest-neighbors sites,
in units of the lattice constant $a$.

In Fig.~\ref{cond4h}, we show the real part of the conductivity
in the undoped case
for  $\lambda = 0$ (undeformed lattice). We can see a dominant feature
formed by the peaks near $\omega = 5$. These peaks correspond to a
charge transfer of a hole from Cu to O ( UHB to NBO
transition).
As one would expect, this absorption feature appears at $\Delta + U_{pd}$
modified by a typical hybridization energy.

As we have discussed in the previous section, the minimum single
particle gap (between UHB and CS) is $\sim 3$.
These transitions are forbidden in our cluster.

In the one-hole-doped system, and for $\lambda = 0$, it is
known that there is a
shift of the Fermi energy towards the CS band. Consequently,
we should see a peak at
low energy. Indeed, this low energy structure may be observed in
Fig.~\ref{cond5h}(full line)
where the first peak at $\omega \approx 2$ corresponds to
transitions from CS to NBO. The structure around $\omega \approx 5$
corresponds to UHB $\rightarrow$ NBO transitions. The upper peak at
$\approx 9$ corresponds to CS $\rightarrow$ LHB transitions.

Now, we examine the optical conductivity in
the region where the BLD is stable.
In Fig.~\ref{cond5h} ( dotted line ), we show the optical conductivity in
the case of $\lambda=1.2$.
The most important change with respect to the undeformed case is a
shift of the peaks to higher energies.
Taking into account the previous analysis, we can associate this change
to the polaronic shift of the single-particle levels.
We also can see  the set of
peaks related to the UHB $\rightarrow$ NBO transitions. These peaks are
approximately in the same place as before. Another shift to higher energies
can be seen in the peak associated with the CS $\rightarrow$ LHB
transitions.
This
peak appears at higher energies because it occurs between
the self-trapped state
and LHB.

\section{Summary and Conclusions}

\indent

In this work, we have examined the self-trapping of an added hole
in the Peierls-Hubbard Model using exact diagonalization in small
clusters.
In the presence of deformation, the hole forms
a local singlet analogous to the Zhang-Rice singlet in the undeformed
lattice but of a more general character, since double occupancy
in Cu sites is not longer negligible.

Our calculation is complementary to inhomogeneous
HF studies\cite{8} and, in general, it confirms the qualitative behavior
found in this approximation. However, there are important differences
between the results obtained by these two approaches. Here we discuss
these differences as well as the connection with other results
in the literature.

Firstly, there is a reduction
of the magnetic moments of the Cu where the doped hole is self-trapped.
However, this effect is much less dramatic than in the HF
calculations.

Secondly, the chemical potential for zero $\lambda$
shifts with doping to a place where spectral weight already
exists in the undoped compound\cite{22,23} in contrast to the HF
results. This is not the case
for non-zero $\lambda$. Such an anomalous behavior
 arises from the lack of lattice dynamics in both our calculation and in
the HF approach.
The observed polaronic shift does not seem strong enough by itself
to explain the observed pinning of the chemical potential.

We have shown that the spectral weight behaves in a similar manner as
reported in
previous studies, where the coupling with the lattice was absent\cite{22,23}.
There is an anomalous transfer of spectral weight but, contrary to those
studies, this transfer is related here to local effects, i.e. it manifests
in local spectral functions.

The peaks in the optical
conductivity are also shifted consistently with
the polaronic shift of the single-particle spectral functions.
This behaviour is qualitatively similar
to the HF plus RPA calculations\cite{17} at electronic energies
in {\em clusters of the same size}.

Thirdly, the small
polaron appears in this study only above a critical value of electron-lattice
coupling. This important difference with respect to HF calculations
can be understood by the fact that these
calculations neglect the translational motion of the correlated
states and the associated kinetic energy.
One expects that $\lambda_c$
will be determined by the competition
between the kinetic energy and the polaronic
effects.

In this exact diagonalization study we have considered a quite small
cluster.
Unfortunately a determination of the size dependence is out of the present
computational possibilities. However, if we assume that $\lambda_c$
scales with the bandwidth of the Zhang-Rice states,
a strong size-dependence
is expected since the effective bandwidth of strongly
correlated systems, for small clusters, decreases with increasing
size\cite{oht}.

The physics close to $\lambda_c$ is highly nontrivial for a finite
value of the ion mass
\cite{der}.
The transition is no longer sharp
but washed out by
quantum fluctuations, and $\lambda_c$ becomes a crossover
value. The renormalization of the
effective mass of quasiparticles is moderate and not exponentially large
like in the large $\lambda$ limit.
In real materials, neither $\lambda$ nor $\lambda_c$
are known  accurately.
The fact that relatively lightly doped cuprates show already
high mobilities coexisting with polaronic signatures,
suggests that the system is close to $\lambda_c$.
A related scenario is that the system forms a more extended polaron with
a BED distortion in a limited region, or fluctuates between both,
localized and extended states\cite{8,15}.

In summary, we have given a complete characterization of the
three-band Peierls-Hubbard model in small clusters in
the adiabatic approximation using exact diagonalization.
We have examined some properties which can not be
properly considered within HF calculations, which have been
so far extensively used to study these systems.
We believe that these effects can be generic to a wide class of materials.

\vspace{1cm}

\large
{\bf Acknowledgments}
\normalsize
\vspace{0.4cm}

A.D. and J.L. are supported in this work by  postdoctoral fellowships
granted by the
European Economic Community. A.D. is also indebted to Fundaci\'on Antorchas
of Argentina for traveling support.
A.G. is grateful to A. Aligia, C. Balseiro, P. Etchegoin, I. Mazin
and R.Zeyher for useful discussions. Special thanks are due to
P. Horsch for a critical reading of the manuscript.
A.G. is supported by a fellowship from the Consejo Nacional
de Investigaciones Cient\'{\i}ficas y T\'{e}cnicas de la Rep\'{u}blica
Argentina.
J.L. is in debt with G. Sawatzky  for many discussions
on how the chemical potential and the spectral weight evolve with
doping  in a strongly correlated system and with F. Barriquand for
a critical reading of the manuscript.
During the realization of this work, J.R. was supported by a
fellowship from the Consejo Nacional de
Investigaciones Cient\'{\i}ficas y T\'{e}cnicas de la Rep\'{u}blica
Argentina.

\vspace{1cm}

*Permanent address: Instituto de F\'isica Rosario (CONICET-UNR) and CIUNR.
Bv. 27 de Febrero 210 Bis. 2000 Rosario, Argentina.

\figure{\label{fig1}
Patterns of displacements of the oxygen ions along the Cu-O bonds for:
(a) Breathing Extended Deformation (BED) and (b) Breathing Local
Deformation
(BLD).}

\figure{\label{fig2}
Ground state energy as a function of $\lambda$ for both, BED and BLD.
The energy is measured with respect to the energy of the undistorted
lattice.}

\figure{\label{fig3}
Equilibrium lattice displacements of the O's  along the Cu-Cu axis as a
function of $\lambda$ for both type of deformations (BED and BLD).}

\figure{\label{fig4}
$\lambda_{c}$ as a function of $U_{d}$.}

\figure{\label{fig5}
Occupancies of the CuO$_4$ cluster as a function of $\lambda$ in the
BLD. We show occupancies in both Cu$_1$ and Cu$_4$ clusters.}

\figure{\label{fig6}
$<S_z^2>$ as a function of $\lambda$ in the BLD. We show $<S_z^2>$ in
Cu$_1$, Cu$_2$ and Cu$_4$ clusters.}

\figure{\label{fig7}
$<Z^{\dagger}Z>$ as a function of $\lambda$ in the BLD.}

\figure{ \label{gstoc}
Single particle spectral function for the half-filled case in (a) Cu and
(b) O sites.  The full and dashed lines
correspond to PES and IPES experiments, respectively. The spectra is
independent of $\lambda$ since for reasonable
 values of $\lambda$ the lattice is undistorted.}

\figure{ \label{g0}
Single particle spectral function  in (a) Cu  and
(b) O sites  for the 1 hole doped
case and $\lambda=0$. The full and dashed lines  correspond to PES and
IPES
experiments, respectively.}

\figure{ \label{g1.2}
Single particle spectral function for
$\lambda=1.2$. The full and dashed lines correspond to PES and IPES
experiments, respectively. We show the spectra in non equivalent Cu's
(a,b,c) and the
total spectra in Cu (d) and O (e).}

\figure{\label{cond4h}
Real part of the conductivity in the undoped case for $\lambda=0$.}

\figure{\label{cond5h}
Real part of the conductivity in the one hole doped case
for $\lambda=0$ (full line) and for $\lambda=1.2$ (dotted line).}


\newpage

\begin{references}

\bibitem{1}
J. G. Bednorz and K. A. Muller, Z. Phys. B{\bf 46}, 189 (1986).

\bibitem{2}
P.W.Anderson, Science, {\bf 235}, 1196 (1987).

\bibitem{3}
F. C. Zhang and T. M. Rice, Phys. Rev. B{\bf37}, 3759 (1988). E. Dagotto
et al, Phys. Rev.B {\bf45}, 10741 (1992).

\bibitem{4}
V. Emery, Phys. Rev. Lett.{\bf58}, 2794 (1987); J.Hirsch,
Phys. Rev.Lett.{\bf 59}, 228 (1987);
C.Varma et al, Solid State Commun. {\bf 62}, 681 (1987).


\bibitem{latt} For a recent overview see
  Proceedings of the Conference {\em Lattice effects in High-Tc
 Superconductors}, Santa Fe, New Mexico, January 1992, edited
 by Y. Bar-Yam et al. (World Scientific, Singapore).



\bibitem{lor2}
J.~Lorenzana and L.~Yu,
 Phys.\ Rev.\ B {\bf 43}, 11474 (1991). J. Lorenzana,
 PhD Thesis (International School for Advanced
 Studies, Trieste, 1992).


\bibitem{lor3}
J.~Lorenzana and L.~Yu,
 Mod.\ Phys.\ Lett. B {\bf 5}, 1515 (1991).

\bibitem{inu}
 M.~Inui and P.~B.~Littlewood,
Phys.~Rev.~B~ {\bf 44}, 4415 (1991).

\bibitem{8}
K.~Yonemitsu, A.R. Bishop, and J.~Lorenzana,
 Phys.\ Rev.\ Lett. {\bf 69}, 965 (1992).

\bibitem{7}
J. B. Grant and A. K. Mc Mahan, Phys. Rev.B {\bf46}, 8440 (1992).

\bibitem{14}
 J.~Lorenzana and L.~Yu,
 Phys.\ Rev.\ Lett. {\bf 70}, 861 (1993).

\bibitem{13}
K.~Yonemitsu, A.R. Bishop, and J.~Lorenzana,
Phys.\ Rev. B {\bf 47}, 8065 (1993).

\bibitem{17}
K.~Yonemitsu, A.R. Bishop, and J.~Lorenzana,
Phys.\ Rev. B {\bf 47}, 12 059 (1993). Idem, Ref. \cite{latt}.

\bibitem{ric}
M. J. Rice and Y. R. Wang,
Phys.\ Rev. B {\bf 36}, 8794 (1987).


\bibitem{9}
Y.~H. Kim, C.~M. Foster, and A.~J. Heeger,
Phys. Scr. T {\bf 27}, 19 (1989).

\bibitem{10}
X.~Wei, C.~Chen, Z.V. Vardeny, C.~Taliani, R.~Zamboni, A.J. Pal, and G.~Ruani,
Physica C {\bf 162-164}, 1109 (1989).

\bibitem{gyu} For a recent review see,
 G. Yu and A. Heeger, Submitted to Rev. Mod. Phys.
This reference discuss the remarkable similarity between the photodoped
and chemically doped experiments.

\bibitem{11}
G.~A. Thomas, D.~H. Rapkine, S-W. Cheong, and L.~F. Schneemeyer,
 Phys.\ Rev.\ B {\bf 47}, 11396 (1993).

\bibitem{15}
J. Zhong, et al, Phys. Rev. Lett.{\bf 69}, 1600 (1992).


\bibitem{16}
J. Wagner, et al, Phys. Rev. B {\bf 43}, 10517 (1991).

\bibitem{P.Hors}
 P. Horsch, et al, Physica C {\bf 162-164}, 783 (1989).


\bibitem{mur}
A. Muramatsu and W. Hanke , Phys. Rev. B {\bf 38}, 878 (1988).

\bibitem{22}
H. Eskes, M.B.J. Meinders, and G. A. Sawatzky,
 Phys.\ Rev.\ Lett. {\bf 67}, 1035 (1991).


\bibitem{23}
M.B.J. Meinders, H. Eskes and G. A. Sawatzky, preprint.


\bibitem{vee} M.A. van Veenendaal, R. Schlatmann, G. A. Sawatzky
and W. A. Groen,  Phys. Rev. B {\bf47}, 446 (1993).


\bibitem{21}
H. J. Schmidt and Y. Kuramoto,
 Phys.\ Rev. B {\bf 42}, 2562 (1990).


\bibitem{oht} Y. Ohta, K. Tsutsui, W. Koshibae, T. Shimozato and
S. Maekawa, Evolution of the in-gap state in High-Tc Cuprates,
preprint.


\bibitem{24}
J. W. Allen et al., Phys. Rev. Lett {\bf 64}, 595 (1990).


\bibitem{19}
E. Gagliano and C.Balseiro, Phys. Rev. B {\bf38}, 11761 (1988).


\bibitem{20}
A. Aligia,
private communication. For the spatial symmetries of the states, see
C. Batista and A.A. Aligia, Solid State Commun. ,{\bf 83}, 419 (1992).
Idem, submitted to Phys. Rev. B.





\bibitem{der}
H. de Raedt, A. Lagendijk, Z. Phys.B {\bf 65}, 43 (1986);
D. Feingberg, S. Ciuchi, F. de Pascuale, Int. J. Mod. Phys. {\bf 4},
1317 (1990); D. Feingberg, J. Ranninger, Phys.Rev. A {\bf 30} (1984);
ibid, {\bf 33} 3466 (1986).



\end{references}
\end{document}